\documentclass{emulateapj}
\usepackage{graphicx}
\usepackage{epstopdf}
\usepackage{amsmath}
\def\be{\begin{equation}}
\def\ee{\end{equation}}
\def\bea{\begin{eqnarray}}
\def\eea{\end{eqnarray}}
\def\ie{{\it i.e. }}
\def\eg{{\it e.g., }}

\journalinfo{Astrophysical Journal, in press}
\slugcomment{}

\shorttitle{Survey design for SED fitting: a FM approach}
\shortauthors{Acquaviva et al.}

\begin{document}

\title{Survey design for Spectral Energy Distribution fitting: \\
a Fisher Matrix approach}

\author{Viviana Acquaviva\altaffilmark{1}, Eric Gawiser\altaffilmark{1}, Steven J. Bickerton\altaffilmark{2}, \\ Norman A. Grogin\altaffilmark{3}, Yicheng Guo\altaffilmark{4}, Seong-Kook Lee\altaffilmark{5}}
\altaffiltext{1}{Department of Physics and Astronomy, Rutgers, The State University of New Jersey, Piscataway, NJ 08854}
\altaffiltext{2}{Department of Astrophysical Sciences, Peyton Hall, Princeton University, Princeton, NJ 08544}
\altaffiltext{3}{Space Telescope Science Institute, 3700 San Martin Drive, Baltimore, MD 21218, USA }
\altaffiltext{4}{Department of Astronomy, University of Massachusetts, Amherst, MA 01003, USA} 
\altaffiltext{5}{School of Physics, Korea Institute for Advanced Study, Seoul, 130-722, Korea}

\begin{abstract}
The spectral energy distribution (SED) of a galaxy contains information on the galaxy's physical properties, and multi-wavelength observations are needed in order to measure these properties via SED fitting. In planning these surveys, optimization of the resources is essential. 
The Fisher Matrix formalism can be used to quickly determine the best possible experimental setup to achieve the desired constraints on the SED fitting parameters. However, because it relies on the assumption of a Gaussian likelihood function, it is in general less accurate than other slower techniques that reconstruct the probability distribution function (PDF) from the direct comparison between models and data.
We compare the uncertainties on SED fitting parameters predicted by the Fisher Matrix to the ones obtained using the more thorough PDF fitting techniques. We use both simulated spectra and real data, and consider a large variety of target galaxies differing in redshift,  mass, age, star formation history, dust content, and wavelength coverage.  We find that the uncertainties reported by the two methods agree within a factor of two in the vast majority ($\sim 90\%$) of cases. If the age determination is uncertain, the top-hat prior in age used in PDF fitting to prevent each galaxy from being older than the Universe needs to be incorporated in the Fisher Matrix, at least approximately, before the two methods can be properly compared. We conclude that the Fisher Matrix is a useful tool for astronomical survey design.
\end{abstract}

\maketitle

\section{Introduction}

The Fisher Information Matrix (FM, \citealt{Fisher35}) is a statistical instrument
of paramount importance in parameter estimation problems. Its function is to {\it predict} the precision with which the parameters of interest can be measured by a given experiment, starting from the expected observational uncertainties on the data. 
The utility of the FM comes from the fact that it can be used to optimize the experimental setup, according to the desired target results, {\it before} actually collecting the data. 

The Cram{\'e}r-Rao inequality \citep{Cramer,Rao} states that the uncertainties predicted by the FM formalism are, in general, optimistic; they set the upper limit for the precision that an experiment can attain. In particular, the FM method is expected to be accurate only if the probability distribution is a Gaussian function of the parameters. Therefore, in order to validate the FM method for a specific problem, it is necessary to compare the predictions obtained via the FM to the results obtained by using other statistical techniques that do not suffer from this limitation. In many fields of Astronomy, the uncertainties predicted by the FM have been shown to closely resemble the ones estimated by fitting the probability distribution function (PDF). 
This is the case, for example, of cosmological parameter estimation from the Cosmic Microwave Background (\eg \citealt{Tegmark1997,Balbi2003}), or distance measurements from Baryon Acoustic Oscillations (\eg \citealt{Seo:2003pu}); see also \eg \citealt{Verde:2009tu} for a review. 

In this paper, we study the applicability of the FM formalism to the problem of fitting the Spectral Energy Distribution (SED) of galaxies. In this case, the planned experiment consists of acquiring multi-wavelength photometry of galaxies, the uncertainties on the projected data are the photometric errors on the flux measurements in each band (including any systematics), and the parameters to be measured are the physical properties of the target galaxies, such as age, stellar mass and dust content. Our objective is to demonstrate, for a variety of cases, that the uncertainties forecasted by the FM are comparable to the ones obtained by running a PDF fitting algorithm on the same simulated data, and therefore the FM predictions can be used as a guideline in survey planning. The advantage of using the Fisher Matrix as opposed to full PDF fitting is that the former method is much faster, allowing one to very quickly predict the expected results for many different experimental setups.

The paper is organized as follows. In Sec. \ref{secI} we review the currently available tools for SED fitting. Sec. \ref{secII} describes the relevant Fisher Information Matrix formalism.  Our main results for the comparison of uncertainties obtained via the FM and MCMC SED fitting are presented in Sec. \ref{secIII}; we consider several cases of galaxies observed with different signal to noise, star formation histories, dust content, and wavelength coverage. 
In Sec. \ref{secIV} we apply GalFish, the Fisher Matrix algorithm we developed, to real data, using a catalog of several hundred galaxies at redshifts $0 < z <  5$ in the Great Observatories Origins Deep Survey South (GOODS-S) field, and we compare the results to those obtained by using GalMC. We provide a worked example of survey optimization in Sec. \ref{secV}. Sec. \ref{secVI} contains a summary of our findings and our conclusions. 

\section{Statistical tools for SED fitting}
\label{secI}
SED fitting is the procedure of comparing models to the observed galaxy SEDs (for reviews, see \citealt{2009NewAR..53...50G,2010Ap&SS.tmp..257W} and references therein). The unknown physical properties of the observed galaxy can be inferred from the known physical properties of the model, once the model that most closely resembles the data (the best-fit model) is found. The best-fit model can be located easily by using $\chi^2$ minimization techniques, \ie building a grid of models and computing the $\chi^2$ for each model in the grid. 
However, the computation of the uncertainties on SED fitting parameters (also called {\it marginalization}) involves solving multi-dimensional integrals, and grid-based techniques are highly inefficient in performing this task. As a result, Markov Chain Monte Carlo techniques have become increasingly popular in the past few years. In MCMC, the parameter space is explored non-uniformly, sampling more frequently where the models resemble the data more closely. As a result, the process is more CPU-efficient, and computing uncertainties amounts to performing sums at the locations sampled by the Markov chain. Examples of MCMC algorithms for SED fitting include \eg \cite{2003MNRAS.343.1145P,2006MNRAS.369..939S,2007A&A...471...71N,2010ApJ...712..833C, 2011arXiv1103.3269S, pirzkal}. In \cite{2011ApJ...737...47A}, we presented our own MCMC fitting algorithm, GalMC, which is publicly available at \small http://www.physics.rutgers.edu/$\sim$vacquaviva/web/GalMC.html. \normalsize
In this paper, we use GalMC to simulate and fit the data, and to compute the marginalized probability distributions for the SED fitting parameters to be compared with the FM predictions.

\section{Fisher Matrix formalism}
\label{secII}
Let us assume that we are observing galaxies in {\it n} photometric bands, so that the data is a vector containing {\it n} fluxes ${\phi^{\rm obs}_{1, .. n}}$, each observed with photometric error ${\sigma^{\rm obs}_k}$ = ${\sigma^{\rm obs}_{1, .. n}}$.  Assuming that our models depend on a set of {\it m} parameters $\vec{\theta} = (\theta_{1, ..  m})$, where we include in this list only the parameters we want to measure, the theoretical fluxes in the same photometric bands are denoted as $\phi^{\rm th}_{1,..n}(\theta_{1,..m})$. For the moment, we assume flat priors in the parameters $\theta_{1, .. m}$, so that the posterior distribution coincides with the likelihood; we return to this issue later in this section.

The likelihood associated to each model is 
\be
{\cal L} = e^{- \chi^2} =  e^{ \sum_{k = 1}^n \frac{(\phi^{\rm obs}_k - \phi^{\rm th}_k(\theta_{1...m}))^2}{(\sigma^{\rm obs}_k)^2}}
\ee

Let $\vec{\theta_{0}}$ be the location of the best-fit model in the parameter space. Around $\vec{\theta_0}$, which is a maximum of the likelihood, we can expand $\ln {\cal L}$ in a Taylor series as:

\be
\ln {\cal L} = \ln {\cal L}(\vec{\theta_0}) + \frac{1}{2} \sum_{ij} (\theta_i - \theta_{i,0}) \frac{\partial^2 \ln \cal L}{\partial \theta_i \partial \theta_j} \Big{|}_{\theta_0}  (\theta_{j,0} - \theta_{j}) + ....
\label{eq:loglike}
\ee

The Fisher Matrix $F_{ij}$, which enters the second term in the above equation, is defined as:
\be
F_{ij} = - \Big{\langle} \frac{\partial^2 \ln \cal L}{\partial \theta_i \partial \theta_j} \Big{\rangle}
\label{eq:F}
\ee

The following relation holds for the marginalized error on parameter $\theta_i$ in terms of the Fisher Matrix:

\be
\label{eq:sigma}
\sigma_{\theta_i} \ge \sqrt{F^{-1}_{ii}}.
\ee

\subsection{Computing the Fisher Matrix}
Equation \eqref{eq:F} offers the most general way to compute the Fisher Matrix. However, in our case this formula can be simplified considerably by using the reasonable assumption that the data obey a Gaussian distribution (in other words, that the likelihood can be written as a Gaussian function of the observable quantities). In this case, it can be shown (\eg \citealt{2009arXiv0906.0664H}) that
\bea
F_{ij} &= &\frac{1}{2} \rm{Tr} [C^{-1} \partial_i C \, C^{-1} \partial_j C \nonumber \\
& + & C^{-1} \partial_i \phi^{\rm th}(\vec{\theta_0}) \, \partial_j (\phi^{\rm th}(\vec{\theta_0}))^T \nonumber \\
& + & C^{-1} \partial_j \phi^{\rm th}(\vec{\theta_0}) \, \partial_i (\phi^{\rm th}(\vec{\theta_0}))^T], 
\eea
where $C$ is the data covariance matrix, and $\partial_i =  \partial/\partial \theta_i$; the trace is computed over the length of the data vector.  
In our case, the photometric errors do not depend on the location of the fiducial model in parameter space, and the first term in the above equation vanishes. The Fisher Matrix formalism is able to describe correlations between measurements in different bands as non-diagonal terms in the covariance matrix. Here we make the simplifying assumptions that they are uncorrelated, and the Fisher Matrix becomes simply
\be
F_{ij} = \sum_{k = 1}^n \frac{1}{(\sigma^{\rm obs}_k)^2} \partial_j \phi^{\rm th}_k(\vec{\theta_0}) \, \partial_i \phi^{\rm th}_k(\vec{\theta_0}).
\ee
As is well-known, the Fisher Matrix does not require the use of any data.

The above formula also allows us to compare the computational advantage of using the Fisher Matrix to predict the uncertainties on SED fitting parameters, with respect to MCMC methods. This comparison is only valid for the specific issue of comparing estimated errors before taking the data, since the Fisher Matrix is not a data fitting technique and, unlike MCMC, does not allow one to find the truly best fitting model and associated uncertainties.

In both cases, almost all of the CPU time is spent running a Stellar Population Synthesis (SPS) code, in our case the faster version of GALAXEV \citep{2003MNRAS.344.1000B,BCprivatecomm} released with GalMC, to compute the theoretical model as a function of the SED fitting parameters. The first derivatives of the model with respect to each parameter can be approximated by the finite differences:
\be
\partial_i \phi^{\rm th}_k(\vec{\theta_0}) = (\phi^{\rm th}_k(\vec{\theta_0} + \Delta {\bf \theta_i}) -  \phi^{\rm th}_k(\vec{\theta_0} - \Delta {\bf \theta_i})) /(2 \Delta \theta_i),
\label{eq:der}
\ee
where $\Delta \theta_i$ is a small increment in the direction of the ${\it i}$-th parameter. Therefore, each derivative requires computing two theoretical models, and the Fisher Matrix requires a total of $2m$ iterations, where $m$ is the number of SED fitting parameters. We have empirically determined in \cite{2011ApJ...737...47A} and \cite{Acquaviva_Evol} that the necessary number of iterations for GalMC to converge is, as a rough rule-of-thumb, $10000m$, larger by a factor of $\simeq 5000$. 
\subsection{Caveats}
The purpose of this paper is to verify whether (or under what circumstances) the ``$\ge$" sign in Eq. \eqref{eq:sigma} becomes effectively a ``$\simeq$" sign. If the uncertainties from the Fisher Matrix are systematically much smaller than the ones found by the slower but more accurate parameter estimation methods which are able to reconstruct the form of the PDF, the Fisher Matrix cannot be used as a guideline in planning experiments. Furthemore, because the Fisher Matrix always assumes a fiducial model, it should only be used in planning experiments and it cannot be a substitute for data fitting techniques such as MCMC.

We expect that differences between the uncertainties obtained by FM and PDF fitting can arise from (at least) three different issues: 

- {\it A non-Gaussian likelihood}. From Eq. \eqref{eq:loglike}, it is clear that if the likelihood is not a Gaussian function of the {\it parameters}, the truncation in the Taylor series is not exact. This is different from the aforementioned assumption of a Gaussian distribution of the data. The latter assumption is well justified at high S/N, while the likelihood is strictly a Gaussian function of the parameters only if the model fluxes are linear functions of the parameters. For a simple star formation history (SFH), this can be expected to be true for stellar mass, which is an overall normalization of the SED, but not, for example, for age or dust content. 

- {\it A displacement between best-fit model and fiducial model}. The Fisher Matrix is computed at some fiducial parameter values, chosen to represent a typical target galaxy. However, due to the photometric error, the data will be in general scattered away from the fiducial values of the fluxes, and the best-fit model might not lie at the fiducial location in parameter space, as shown in Fig. \ref{fig:SED}. If the shape of the posterior varies significantly with the location in parameter space, the size of the uncertainties predicted by the FM at the fiducial values may not reflect those at the location of the best-fit model (and as a consequence, those found by PDF fitting). In the case of real data, this issue can be further worsened by template incompleteness, although this is not a specific drawback of the FM formalism but rather a consequence of assuming a fiducial model.

- {\it The impact of priors used in PDF fitting}. In the expression for the Fisher Matrix, we have implicitly assumed flat priors in the parameters. While this is not in general necessary (and, for example, using a uniform prior in $\ln$(age) or $\ln$(mass) is straightforward), it is not possible to use top-hat priors (define a range) in the FM formalism. This issue is relevant especially for the upper bound in age corresponding to the age of the Universe. Such prior cannot be taken away from the MCMC run used for PDF fitting, not only because it is an important piece of knowledge that we want to incorporate in the sampling process, but also because theoretical models are only available in a limited age range. However, there are ways to simulate priors of similar effect in the FM, as explained in Sec. \ref{sec:lm}. 

In the next sections we present the results of our careful comparison of the uncertainties predicted by FM and MCMC SED fitting, commenting on each of these issues.

\section{Fisher Matrix vs. MCMC - Simulated data}
 \label{secIII}
 
\begin{figure}[t!]
\centering
 \includegraphics[width=\linewidth]{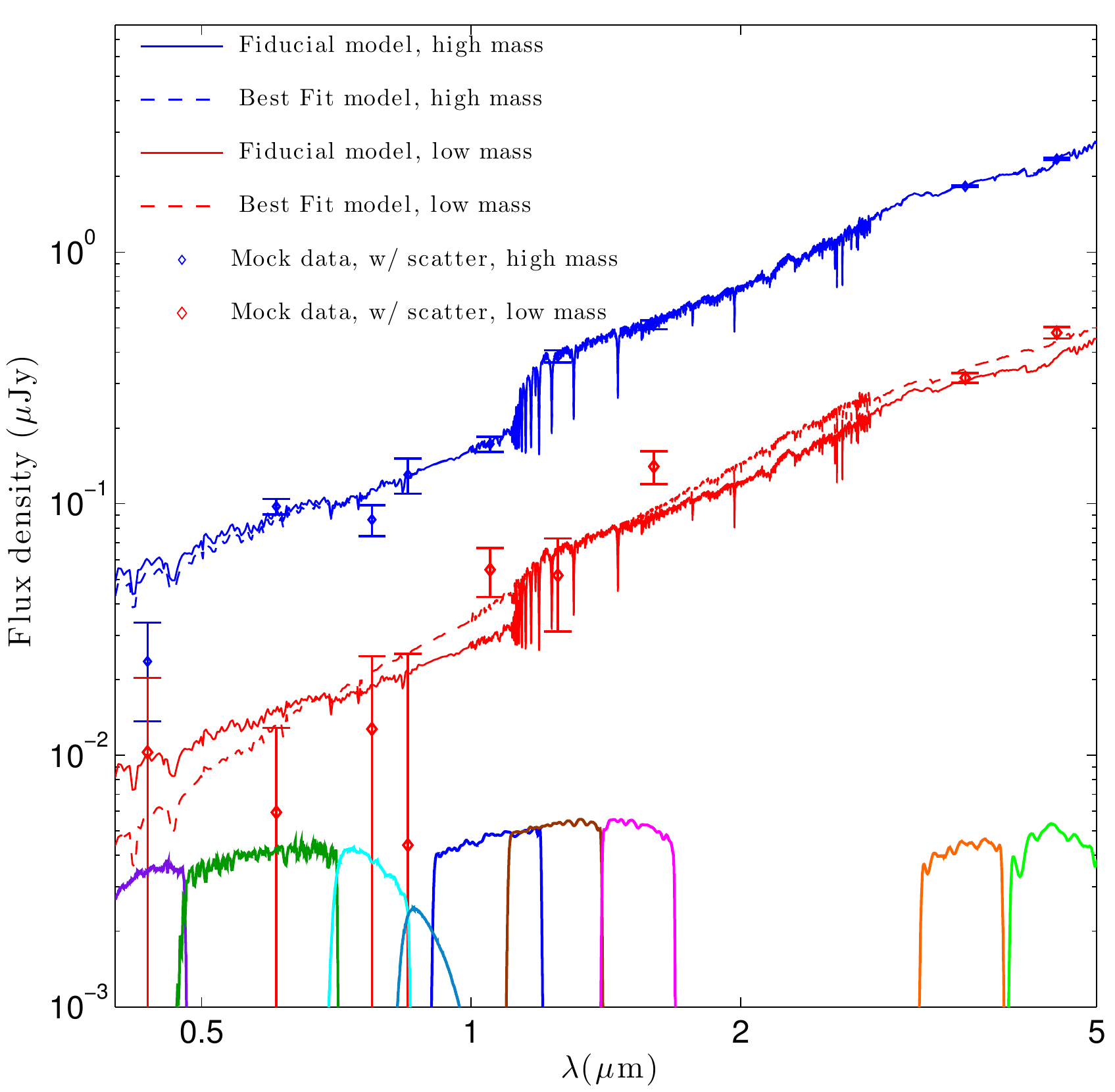}
 \caption{Mock data, fiducial and best-fit SED for one high-mass and one low-mass galaxy at z = 2 considered in the analysis, showing the signal-to-noise corresponding to the limiting magnitudes listed in Table \ref{table:mag}. The mock data points are obtained by adding a Gaussian scatter of $1-\sigma$ amplitude equal to the photometric error to the model fluxes; the best-fit model comes from SED fitting using GalMC. For the low-mass galaxy the displacement between the best-fit and the fiducial model is more pronounced, since the signal to noise ratio is lower and the data have less constraining power. The filter transmission curves (multiplied by 0.001) are also plotted for reference.}
 \label{fig:SED}
 \end{figure}

We conduct our analysis on different sets of reference galaxies. The assumed wavelength coverage and limiting magnitudes (summarized in Table \ref{table:mag}) are inspired by Fig. 3 of \cite{2011arXiv1105.3753G}, which corresponds to the ``Wide" program of the CANDELS galaxy survey \citep{2011arXiv1105.3753G,2011arXiv1105.3754K}, supplemented by photometry in the Spitzer IRAC 3.6 and 4.5 $\mu$m bands from GOODS \citep{Dickinson:2002va}. Therefore, the results found here represent what is achievable, using public data only, for a galaxy in GOODS-S. We do not include any additional calibration uncertainty. We consider 1 Gyr old galaxies at different redshifts (z = 1, 2, and 3) with high ($M_{*} \simeq 9 \times 10^9 M_\odot $) and low ($M_{*} \simeq 1.5 \times 10^9 M_\odot$) stellar mass. We marginalize over age, total mass M$_{\rm tot}$ (the integral of the star formation history over the age of the galaxy, which is the input of SPS codes), and dust content parametrized by the excess color E(B-V). Fig. \ref{fig:SED} shows the simulated data (including a scatter of width equal to the photometric error), the fiducial model SED, and the best-fit SED for the high-mass and low-mass reference galaxy at z = 2, together with the filters' transmission curves.

For all the galaxies described below, we compute the derivatives of the SED (convolved with the 9 filter curves) using Eq. \eqref{eq:der}. The derivative with respect to the total mass can also be computed analytically since the SED depends linearly on the total mass, and we verified that the derivatives obtained with the two methods agree.
The step size is chosen to be 5\% of the reference parameter value, with the exception of dust-free galaxies, where we used values of E(B-V) = 0.01 and -0.01 (which is unphysical, and only used in order to correctly evaluate double-sided derivatives). We checked that the step size is unimportant by choosing different values and verifying that the derivatives do not change. 

The MCMC analysis in GalMC is conducted using a flat prior in $\ln$(age) and $\ln$(mass). In the Fisher Matrix, we used both a flat prior in age and $\ln$(age), and mass and $\ln$(mass), finding no or very little (at the percent level) dependence on this choice of prior, which won't be further discussed.

 \begin{figure}
 \centering
 \includegraphics[width=\linewidth]{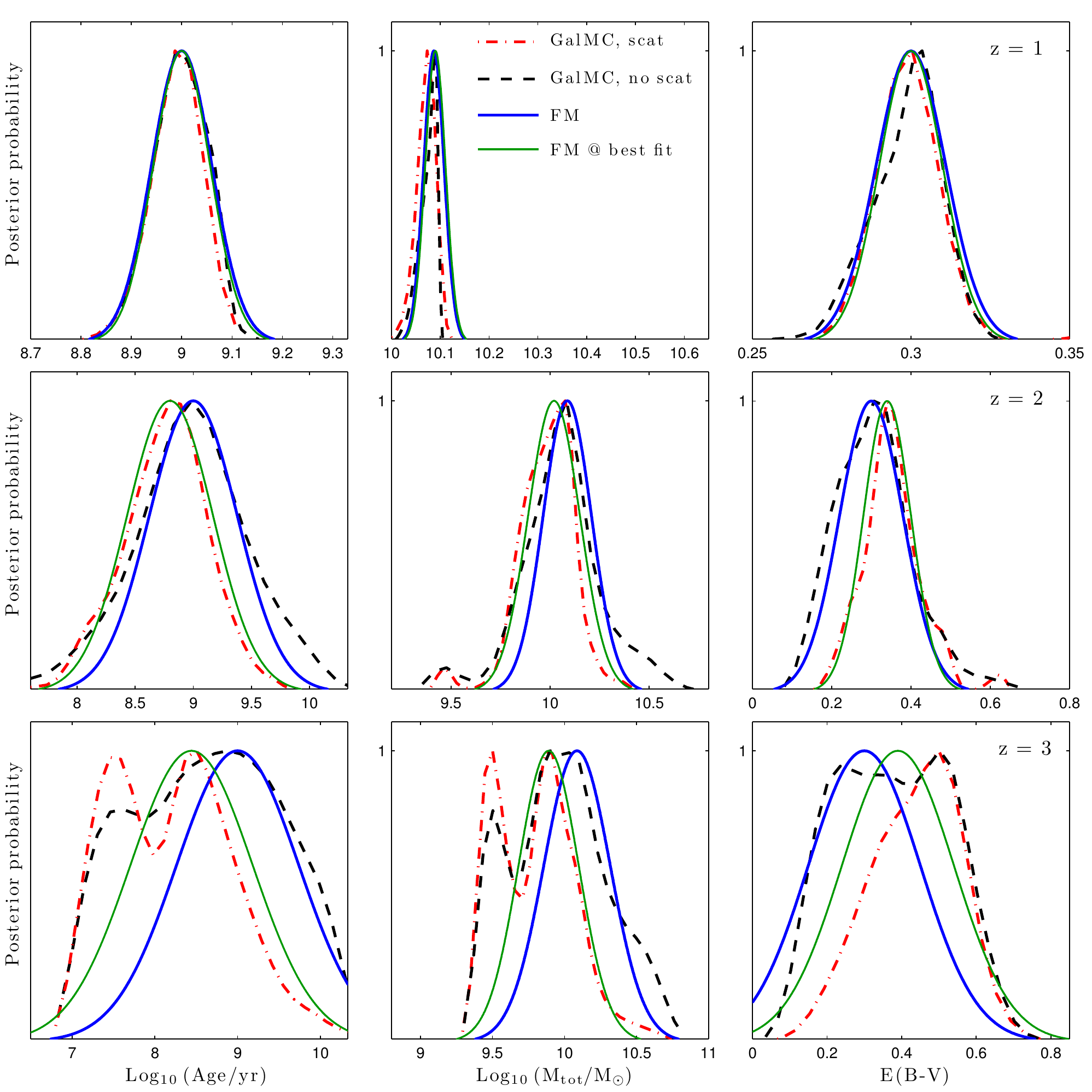}
\caption{{\bf High-mass galaxies.} Comparison of uncertainties on SED parameters from the Fisher Matrix prediction (FM, solid lines) and GalMC (dashed lines) for high-mass galaxies ($M_* \simeq 8\times 10^9 M_\odot$). Each row corresponds to a different redshift (z = 1, 2, and 3). 
The FM is computed at the fiducial value (blue thicker line) and at the best-fit model obtained by GalMC when scatter is added to the data (thin green line). The PDF from GalMC is shown for simulated data without (dashed line) and with (dot-dashed line) scatter. The estimates from FM and GalMC are in good agreement in all cases; see text for more details.}
\label{fig:dusty_mas}
\end{figure}

 \subsection{High-mass galaxies}
The first set of galaxies we consider are massive galaxies ($M_{*} \simeq 9 \times 10^9 M_\odot $) at the three reference redshifts z = 1, 2 and 3, characterized by constant star formation (CSF) history and a moderate amount of dust (E(B-V) = 0.3).  We begin by simulating data with uncertainties corresponding to the magnitude limits in Table \ref{table:mag}, but not adding scatter to the simulated data points. In this case, we are testing the first of the ``caveats" listed in the previous section; we expect the constraints coming from GalMC to be equal to or  broader than those predicted by the FM, as a result of a non-Gaussian PDF which cannot be captured by the FM formalism.

Our results are shown in Fig. \ref{fig:dusty_mas}. The agreement between GalMC and FM is very good for z = 1 and z = 2 galaxies, the main difference being some non-Gaussian tails coming from the MCMC analysis. In the case of the galaxy at z = 3, the FM is obviously unable to capture the bi-modality of the posterior, and the FM moderately underestimates the uncertainties (at the 20\% level). This behavior is probably a consequence of the lower S/N ratio due to the increase in the luminosity distance at higher redshift. 

Next, we introduce a Gaussian scatter in the mock data, of amplitude equal to the photometric error on each data point, and use GalMC to fit the new simulated data. This setting corresponds to what we can expect with real data. Strictly speaking, the FM is expected to represent the PDF at the location of the best-fit model, so finding a difference in the width of the distribution coming from MCMC, which uses the simulated data, and the FM, which only uses the theoretical models at the fiducial value, is not in itself worrisome. However, if the PDF at the best-fit model is very different from the PDF at the fiducial model, the utility of the FM to predict uncertainties before performing an experiment is diminished, since in that case knowledge of the best-fit model is not available.

The results are also shown in Fig. \ref{fig:dusty_mas}. We find a good agreement (within 20\%) between the width of the distribution obtained by GalMC using the data including scatter, and the FM previously computed at the fiducial model values. To verify that the disagreement (and in particular, the smaller width of the dust distribution reconstructed by GalMC at z = 2)  indeed comes from the displacement of the best fit model with respect to the fiducial one, we also compute the uncertainties coming from the FM method at the location of the best-fit model. In this case the uncertainties from the FM are correctly equal to or smaller than the uncertainties coming from GalMC. In the subsequent analysis, we only compare the FM uncertainties evaluated at the fiducial model with the distribution obtained by GalMC by fitting simulated data including scatter, since this is the relevant case for survey planning.

  \begin{figure}[t]
  \centering
 \includegraphics[width=\linewidth]{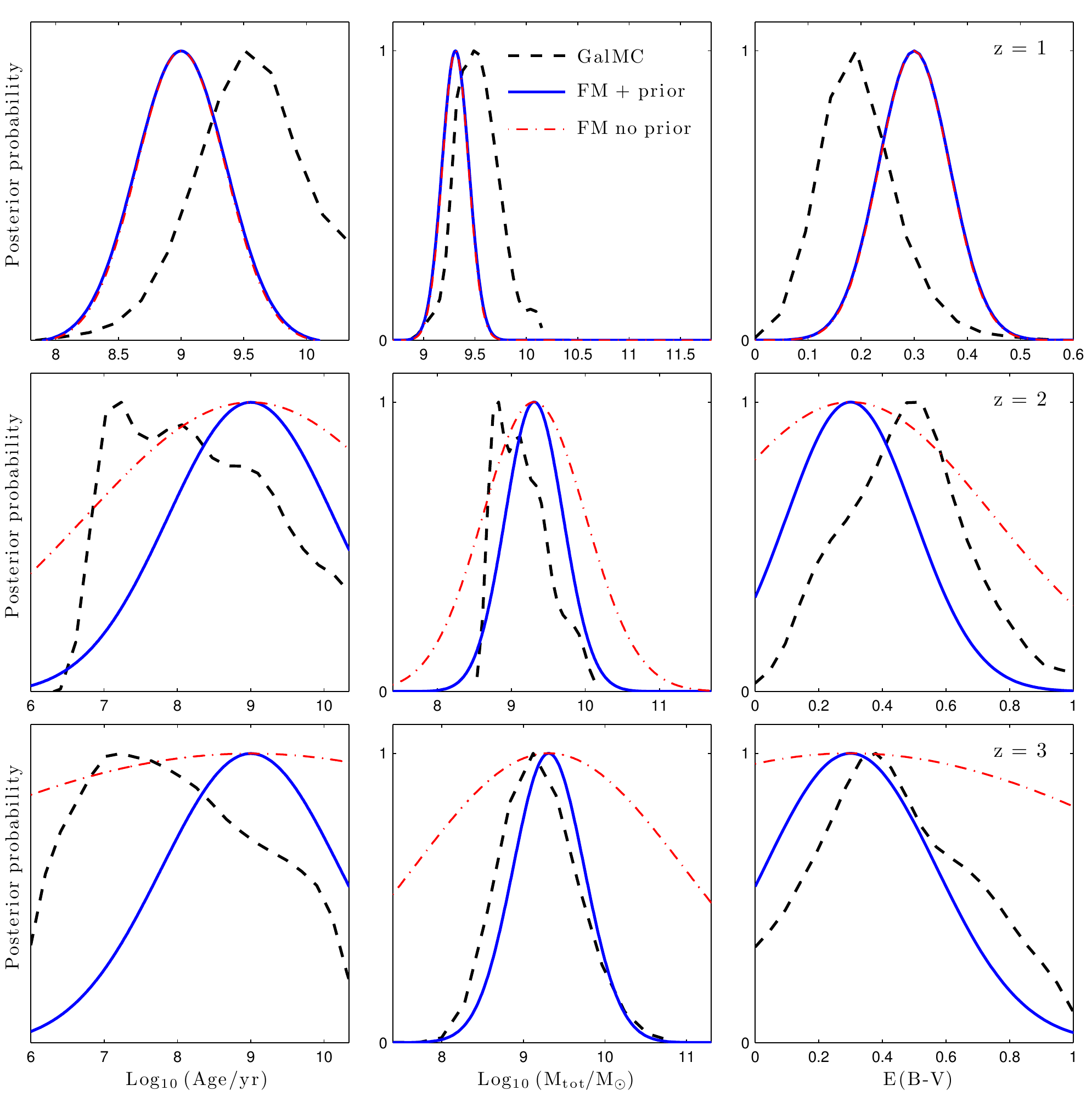}
 \caption{{\bf Low-mass galaxies.} Comparison of uncertainties on SED parameters from the Fisher Matrix prediction at the fiducial model, including a broad Gaussian prior on age (blue solid lines) or not (red dashed-dotted lines) and GalMC (black dashed line) for low-mass galaxies ($M_* \simeq 1.5 \times 10^9 M_\odot$). Mock data include a Gaussian scatter. Each row corresponds to a different redshift (z = 1, 2, and 3). 
The MCMC predictions include a top-hat prior on age. At redshift z = 1 (first row), the S/N is still good enough that the prior is inessential and the FM predictions match the results of GalMC, although the uncertainty in mass is underestimated by the FM by about 50\% . At redshift z = 2 and z = 3 (bottom two rows), the prior used in MCMC needs to be incorporated in the FM, which would otherwise overestimate the uncertainties. When the prior is taken into account, the uncertainties from the two methods are in good agreement for all parameters.} 
\label{fig:dusty}
 \end{figure}

 \subsection{Low-mass galaxies}
 \label{sec:lm}

 The next set of galaxies that we consider are similar in SFH and dust content to the ones previously discussed, but have lower stellar mass of $M_{*} \simeq 1.5 \times 10^9 M_\odot $, and consequently, lower signal to noise. We expect SED fitting to be more challenging in this case, since the displacement between best-fit and fiducial model will in general be larger, as illustrated in Fig. \ref{fig:SED}, potentially causing a sizable disagreement between the width of the PDF at the fiducial model evaluated by the FM and the width of the ``real" PDF reconstructed by GalMC.
 
The results are shown in Fig. \ref{fig:dusty}. At redshift z = 1, 
the FM predictions are similar to the results from GalMC, although the uncertainty in the mass is underestimated by the FM by about 50\%. The anticipated difference between fiducial model and best-fit model causes the observed displacement of the peaks in the two curves; however, the widths of the two distributions, which represent the size of the uncertainties, are very similar.  At higher redshift, though, the FM appears to overestimate the attainable precision on all parameters. In this case, the difference is caused by the top-hat prior on age (between a minimum value of $10^6$ yr and a maximum value of $14$ Gyr) used by GalMC. To properly compare the two methods, one would need to use the same prior in the FM formalism. However, it is not possible to include a flat ``range" in the FM, while it is easy to implement a Gaussian prior in any parameter just by adding it to the corresponding diagonal terms of the Fisher Matrix. Therefore, we add a broad Gaussian prior of width $\Delta \log_{10}$(age/yr) = 1.6 when computing the FM; we verified that the exact width of the prior (between 1.3 and 1.75) is not essential.  This simple addition is able to reproduce the main effect of the top-hat prior in age used by GalMC, and when the prior is enforced, the uncertainties predicted by the FM or found by GalMC are in good agreement.
As an empirical prescription for FM calculation, we propose in general to run the FM both with and without this prior. If no dependence on the prior is found, such as, for example, for the low-mass galaxies at z = 1, or for all the high-mass galaxies previously discussed (not shown), one can be confident in the FM results. In general, this will be the case when the tails of the distribution predicted by the FM are far away from the edges of the top-hat priors set in GalMC. On the contrary, if the priors set in GalMC intersect the distribution predicted by the FM at high likelihood values (near the center, rather than near the tails), one should keep in mind that these uncertainties might be grossly overestimated, and possibly use a Gaussian prior in the FM in order to capture the main effect of the top-hat prior used by MCMC.
GalMC also uses a top-hat prior in E(B-V) between 0 and 1. While we didn't find this prior to be of importance in this case, since the posterior distribution intersects these extrema at low values of the likelihood, it might be useful in the general case to test that the FM prediction does not change significantly if a Gaussian prior in E(B-V) is used.
A special case is represented by galaxies with very high (E(B-V) $\simeq$ 1) or very low (E(B-V) $\simeq$ 0) dust content, which we discuss in detail in the next subsection.

\begin{figure}[t]
\centering
 \includegraphics[width=\linewidth]{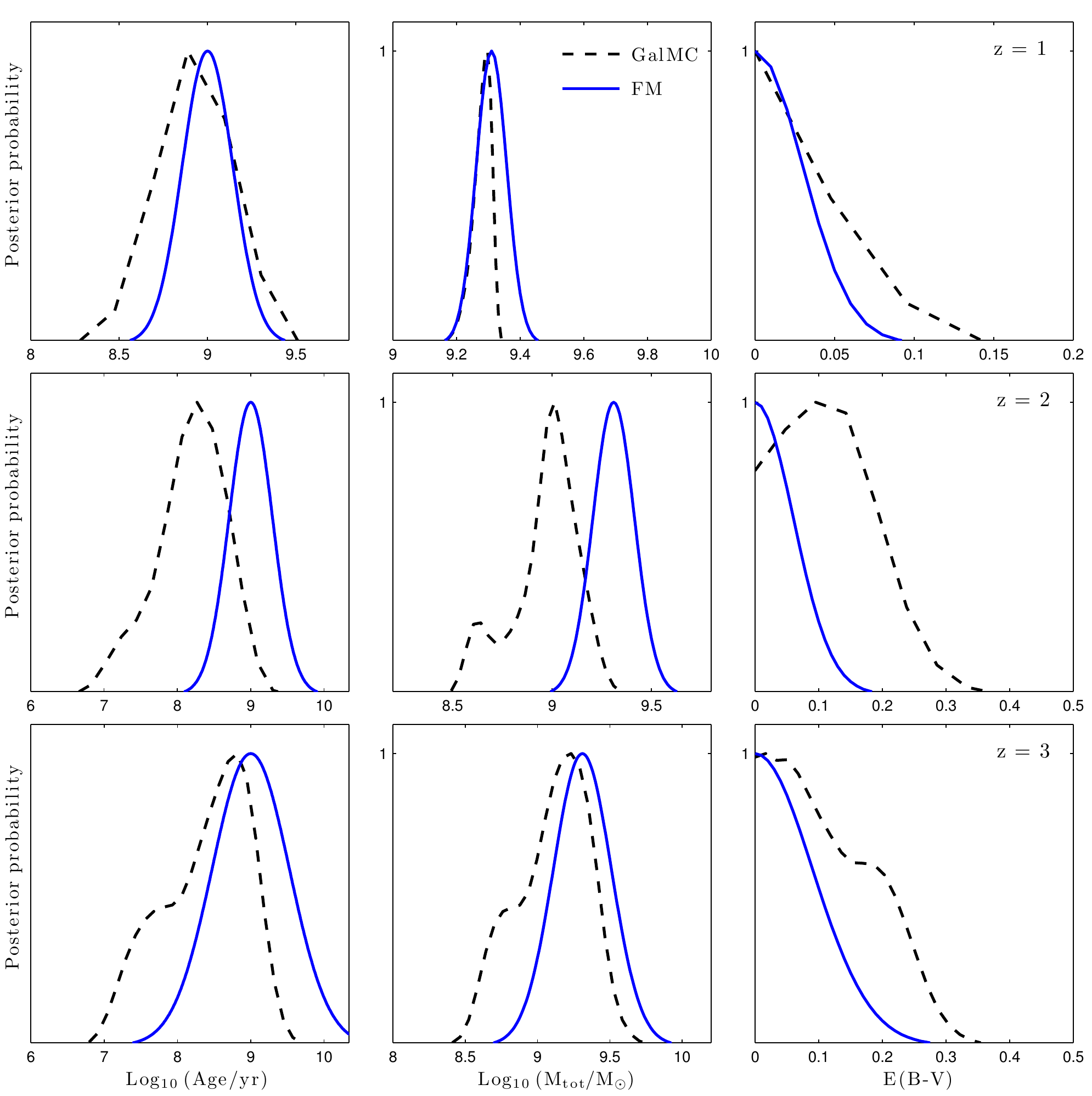}
 \caption{{\bf Dust-free galaxies.} Comparison of uncertainties on SED parameters from the Fisher Matrix prediction at the fiducial model (blue solid line) and GalMC (black dashed line) for dust-free low-mass galaxies ($M_* \simeq 1.5 \times 10^9 M_\odot$). Mock data include a Gaussian scatter. Each row corresponds to a different redshift (z = 1, 2, and 3). 
Since dust-free galaxies have higher signal to noise in the rest-frame UV-optical than their dusty counterparts, the constraints are significantly better than those from Fig. \ref{fig:dusty}, and the prior on age is unimportant. The slight underestimation of uncertainties from FM (order of 30\%) in this case might be due to the constraint E(B-V) $>$ 0 used by GalMC, which cannot be reproduced by the FM. However, this effect is mild and well within the uncertainties observed in other cases.} 
\label{fig:nodust}
 \end{figure}

 \subsection{Low-mass galaxies, dust-free}
 Dust-free galaxies present a possible challenge for the Fisher Matrix vs MCMC comparison, because the prior that forbids negative values of E(B-V), used in MCMC, cannot be incorporated in the FM formalism. The latter only allows one to introduce priors centered at the fiducial model, and any Gaussian prior centered at E(B-V) = 0 and wide enough not to introduce a spurious extra weight at low-dust values does not have any effect on the shape of the Fisher Matrix. The effect of this difference would be an underestimation of the width of the distribution by the FM. In fact, all models with E(B-V) $< 0$ are rejected during MCMC sampling, so that more models with non-zero amounts of dust become part of the Markov chain during the random walk, making the posterior distribution (which is proportional to the density of sampled points) wider than it would be in absence of this constraint. While we observe this effect in our comparison in Fig. \ref{fig:nodust}, the difference in the distribution recovered by GalMC and the one predicted by the FM is still moderate (around 30\%), even for these low-mass galaxies. We notice in particular that a large part of the visually striking difference in the dust distribution at z = 2  is attributable to the median value for dust being displaced from the fiducial model by about $\Delta$E(B-V) = 0.1, while the width of the PDFs from FM and GalMC is indeed similar. Unlike the case of their dusty counterparts, in dust-free galaxies the rest-frame UV and optical flux is not suppressed, enabling a better determination of age and making the use of a prior in age unnecessary.

 \begin{figure}[t]
\centering
 \includegraphics[width=\linewidth]{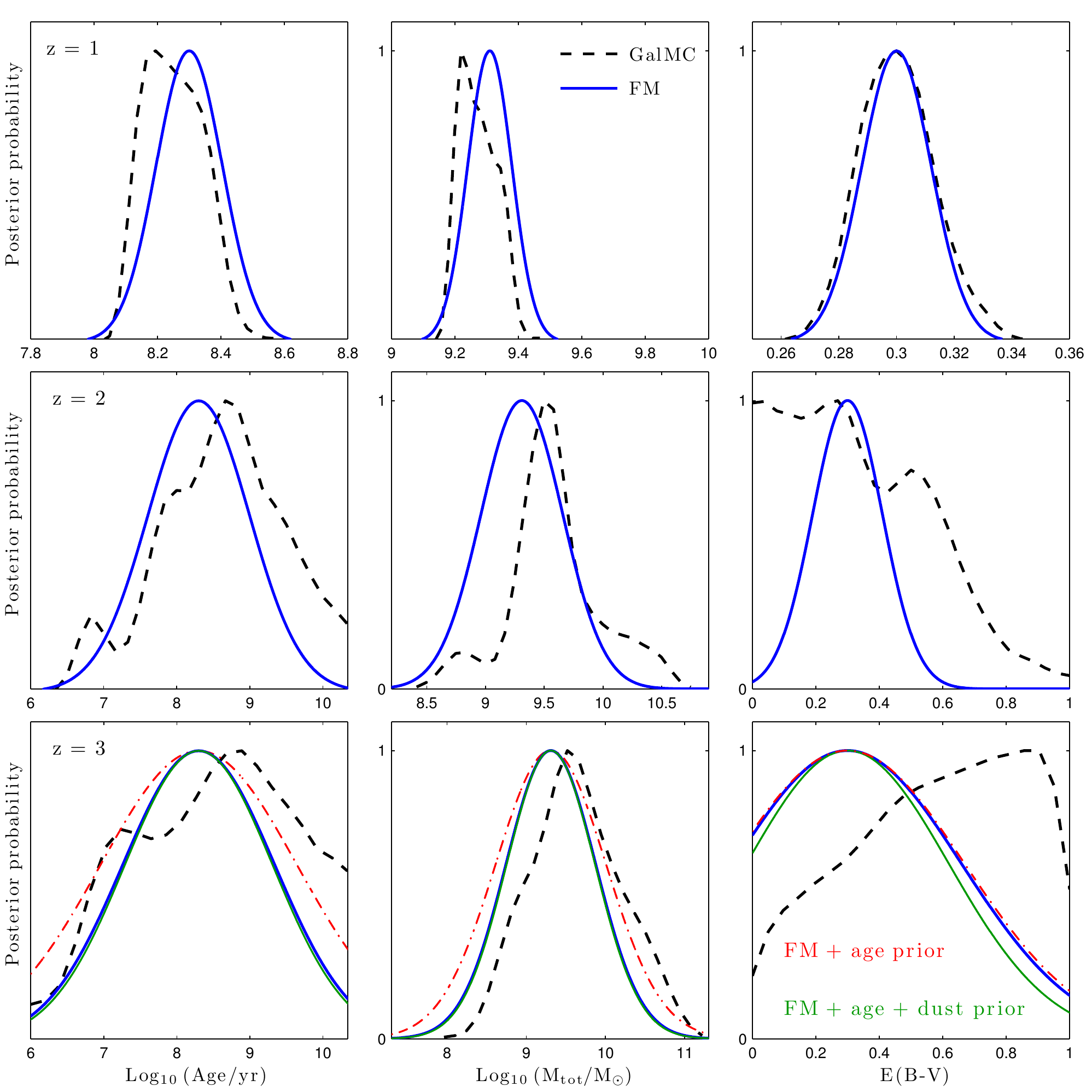}
 \caption{{\bf Starburst galaxies.} Comparison of uncertainties on SED parameters from the Fisher Matrix prediction at the fiducial model (blue solid line) and GalMC (black dashed line) for starburst low-mass galaxies ($M_* \simeq 1.5 \times 10^9 M_\odot$). Mock data include a Gaussian scatter. Each row corresponds to a different redshift (z = 1, 2, and 3). For these galaxies the parameter ``age", identified with the time from the starburst, is 200 Myr. The results obtained by the two methods are in good agreement. For the z = 3 galaxy we also show the effect of a broad Gaussian prior in age (red dashed-dotted line) and in E(B-V) (green thin solid line). The former has a moderate impact on the prediction for age, while the latter is basically negligible. The priors have no effect for z = 1 and z = 2 galaxies, as expected (see discussion in the text).}
 \label{fig:ssp}
 \end{figure}

 \subsection{Low-mass galaxies, SSP}
 
To verify whether the star formation history has an impact on the FM vs PDF fitting comparison, we used the two methods to estimate the uncertainties on SED fitting parameters for low-mass starburst galaxies, often referred to as ``simple stellar populations" (SSP). In this case, we use the terminology ``age" slightly inappropriately, since we are basically only sensitive to the time elapsed since the instantaneous burst of star formation happened. In this scenario, the rest-frame UV and optical radiation coming from massive stars decreases rapidly with time, since these stars have shorter life spans, and these energetic photons are not replenished by ongoing star formation as in the CSF case. One cannot hope to recover the galaxy properties well for a burst that happened 1 Gyr before observation (\ie, at age = 1 Gyr). Therefore, we take as reference cases galaxies where the time since the starburst is 200 Myr. The results of our comparison, shown in Fig. \ref{fig:ssp}, confirm that the agreement between the width of the distributions predicted by the FM and computed by GalMC is good (within 20\% after the prior on age is properly taken into account for the higher-redshift objects, with the exception of the dust content for the z = 2 galaxy, which is underestimated by a factor two). For z = 3 galaxies, we noticed that the prior at E(B-V) = 0 used by GalMC ``cuts" the FM distribution at a high likelihood value. To possibly reproduce this behavior in the FM formalism, similarly to what we did with age, we introduced an additional wide ($\Delta$E(B-V) = 0.8) Gaussian prior in the dust distribution, centered at the fiducial model (E(B-V) = 0.3). We found however that the effect of such prior is negligible (bottom-right panel of Fig. \ref{fig:ssp}).
 
  \begin{figure}
\centering{ \includegraphics[width=\linewidth]{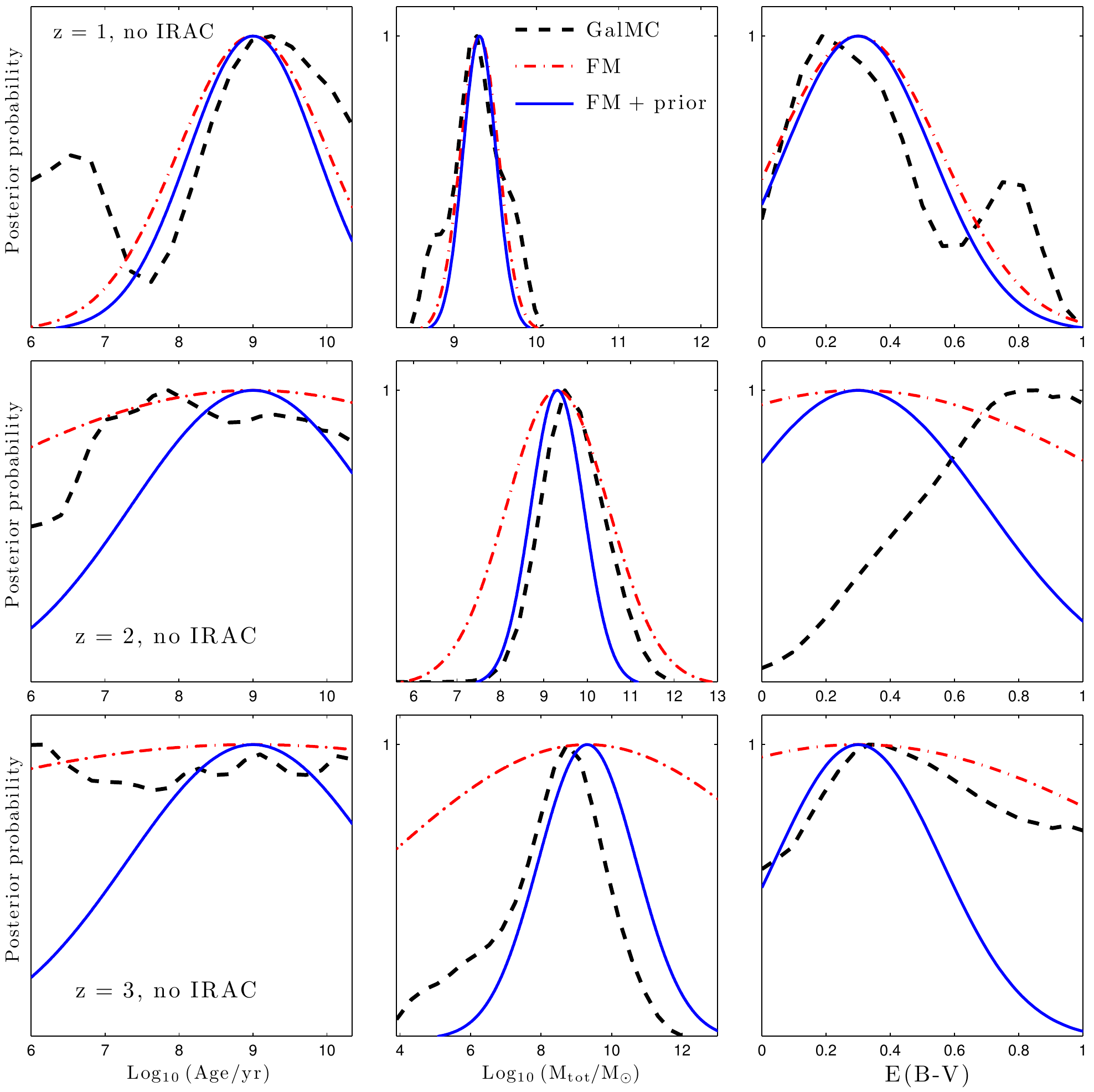}}
\caption{{\bf No IRAC coverage.} Comparison of uncertainties on SED parameters from the Fisher Matrix (FM) prediction at the fiducial model and GalMC for low-mass galaxies ($M_* \simeq 1.5 \times 10^9 M_\odot$), when the IRAC 3.6 and 4.5 $\mu$m bands are not available. Mock data include a Gaussian scatter. Each row corresponds to a different redshift (z = 1, 2, and 3). As in previous cases, the top-hat prior on Age used in GalMC must be taken into account in computing the FM predictions, at least at z = 2 and z = 3 (bottom rows); the width of the PDFs obtained from the two methods are indeed comparable. As expected, the constraints on SED fitting parameters become poorer for increasing redshift, since the lack of rest-frame NIR information causes a dramatic increase in the uncertainties.}
\label{fig:noIRAC}
\end{figure}

 \subsection{Low-mass galaxies, no IRAC coverage}
  
 As a final test case for our comparison, we considered low-mass galaxies for which no IRAC coverage is available. It is well known that the rest-frame near-infrared spectrum is a good tracer of stellar mass, so we expected that determining the galaxy mass would be extremely challenging, and we aimed to verify that the FM formalism would be able to capture this phenomenon. Our results are shown in Fig. \ref{fig:noIRAC}. At redshift z = 1, the handle offered by the near-IR YJH coverage is still able to guarantee a measurement of stellar mass within a factor of two, and the prior in age has a very moderate effect on the shape of the FM. At higher redshift, the complete lack of constraints on the SED beyond rest-frame optical wavelengths causes a dramatic increase in the uncertainties in all SED fitting parameters. In these cases, imposing a prior in age in the FM formalism, as discussed in Sec. \ref{sec:lm}, becomes important. After the prior has been incorporated, we find once again good agreement between the prediction of the FM and GalMC's results, although, as expected, the parameters are basically unconstrained within their ranges. 
 
 \section{Fisher Matrix vs. MCMC: Application to data from GOODS-S}
 \label{secIV}
 The comparison between the results of the Fisher Matrix approximation and SED fitting has so far only been applied to a set of simulated data. However, when the MCMC algorithm is used on real data, we can expect further differences to be introduced by template incompleteness (the fact that real galaxy spectra are more complicated than the models we used to describe them) and by the residual effect of priors, as noticed above. To investigate this issue, we use a sample of 640 galaxies at $0 < z < 5$ with spectroscopic redshifts in the GOODS-S field, which are part of the catalog presented in \cite{Dahlen2010}.  Each SED contains twelve data points, VLT/VIMOS (U band), HST/ACS (F435W, F606W, F775W, and F850LP bands), VLT/ISAAC (J, H, and Ks bands), and the four Spitzer/IRAC channels (3.6, 4.5, 5.8, and 8.0$\mu$m). A $5\%$ flux error in the U, optical and NIR bands, and a 10$\%$ error in the IRAC bands is added in quadrature to the photometric error to account for calibration uncertainty and template incompleteness.

  \begin{figure}
 \centering
 \includegraphics[width=\columnwidth]{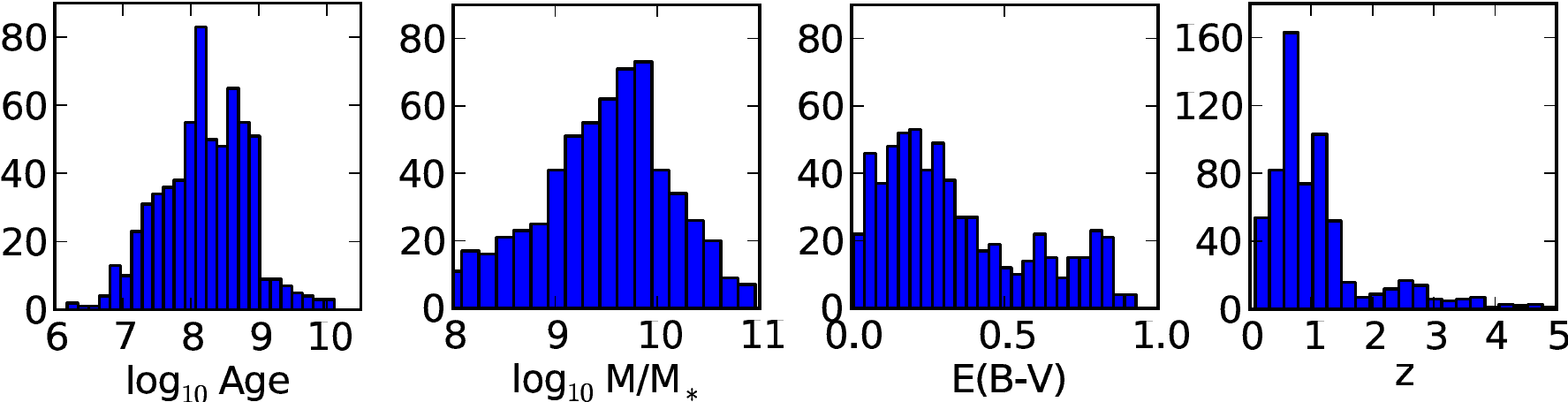}
\caption{{\bf Physical properties of the test sample.} The distributions of mean ages, stellar masses, and E(B-V) values for the set of 485 galaxies in GOODS-S used in Sec. \ref{secIV} reveal the heterogeneity of the sample. The redshift distribution is also shown in the rightmost panel.} 
\label{Fig:params}
\end{figure}

\begin{figure}
 \centering
 \includegraphics[width=\linewidth]{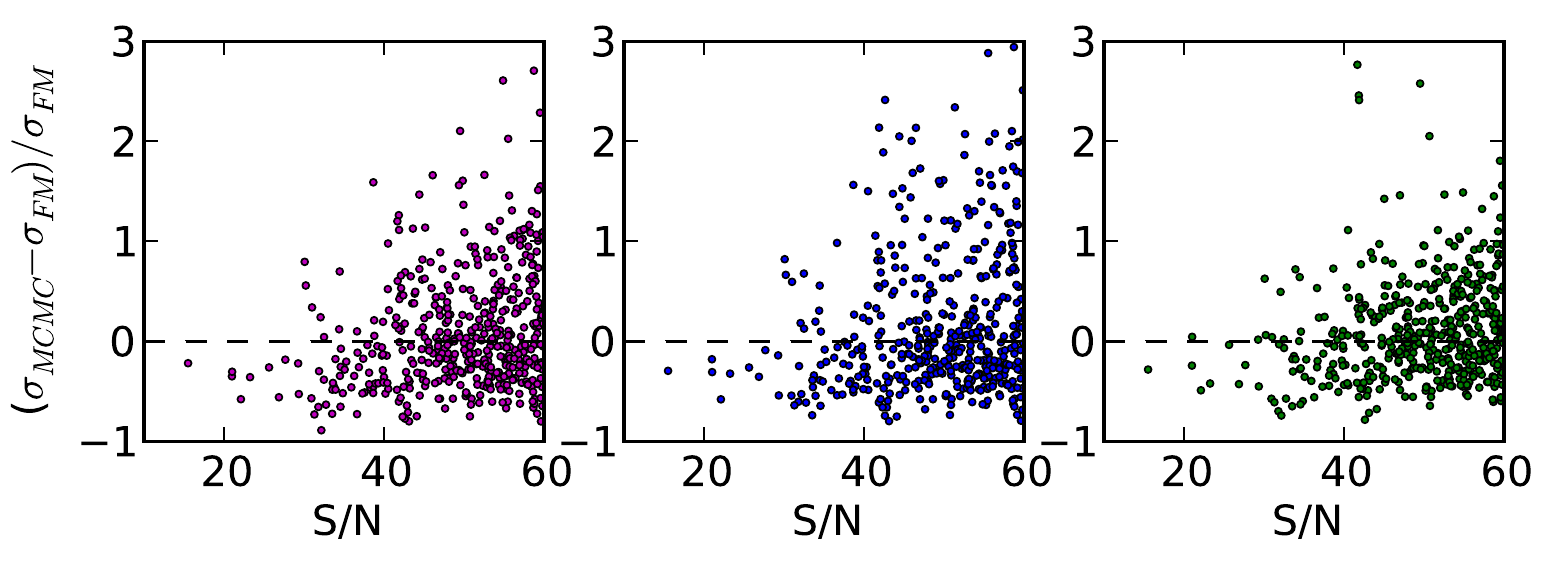}
  \includegraphics[width=\linewidth]{Cutout_All_SN.pdf}
   \includegraphics[width=\linewidth]{Cutout_All_SN.pdf}
\caption{{\bf GalMC vs GalFish for galaxies in GOODS-S.} The figure shows the relative difference in the errors on the SED fitting parameters from the two methods, as a function of $\chi^2$ (first row), signal-to-noise ratio in the SED (second row), and star formation rate (third row). We do not observe any significant correlation between these properties and the ability of GalFish to match the uncertainties of the MCMC analysis.} 
\label{Fig:FMGalMC}
\end{figure}

For each of these galaxies, we use GalMC to perform MCMC SED fitting and select the objects for which the chains pass all the convergence tests, and the reduced $\chi^2$ value is less than 10. In fact, SED fits with very high reduced $\chi^2$ value indicate a strong template incompleteness problem, and the MCMC uncertainties are likely to be significantly underestimated in these cases. This procedure selects 485 objects, characterized by a broad range of physical properties; the distribution of the mean expectation values of SED fitting parameters and the redshifts for the sample are illustrated in Fig. \ref{Fig:params}. We proceed to compute the uncertainties on the SED fitting parameters predicted by our Fisher Matrix algorithm, GalFish, for this set of galaxies. In each case, we use as reference spectrum (around which the derivatives are computed) the one of the model characterized by the mean expectation values of the parameters for that galaxy.
The results of our test are shown in Fig. \ref{Fig:FMGalMC}, which summarizes the differences between the uncertainties found by the MCMC and the ones predicted by the FM, expressed by the quantity $(\sigma_{\rm MCMC} - \sigma_{\rm FM})/\sigma_{\rm FM}$ for each of the parameters.

In the case of age, the average value of the relative difference is 10\%, meaning that if the uncertainty on age predicted by FM is 20\%, the uncertainty found by MCMC would be $18-22\%$. An useful way to summarize the overall behavior of the test is the fraction of objects for which the absolute value of the relative difference is less than 0.5 or less than 1.0, which are the 68 and 91\% of the sample. One possible caveat associated to this statistic is that objects for which $\sigma_{\rm MCMC} << \sigma_{\rm FM}$ would still satisfy the condition $|(\sigma_{\rm MCMC} - \sigma_{\rm FM})/\sigma_{\rm FM}| <$ 1.0, despite the disagreement between the uncertainties predicted by the two methods. However, such occurrences (which amount to about 10\% of the objects that pass the test) are much more likely to be due to MCMC underestimating the uncertainties because of template incompleteness, rather than to the failure of the Fisher Matrix formalism. In fact, if we alleviate the template incompleteness problem by considering only the subset of SEDs for which $\chi^2_r < 3$, these occurrences basically disappear. We also note that the median value of the relative difference is slightly negative (- 5\%), while one would generally expect this quantity to be positive, since the FM should give an optimistic evaluation of achievable uncertainties. However, for this test we have assumed only an extremely broad Gaussian prior in age, which basically has no effect on the parameters, while the MCMC uses a top-hat prior on this quantity. Therefore, this result is not problematic.

  \begin{figure*}
 \centering
 \includegraphics[height = 5.5cm]{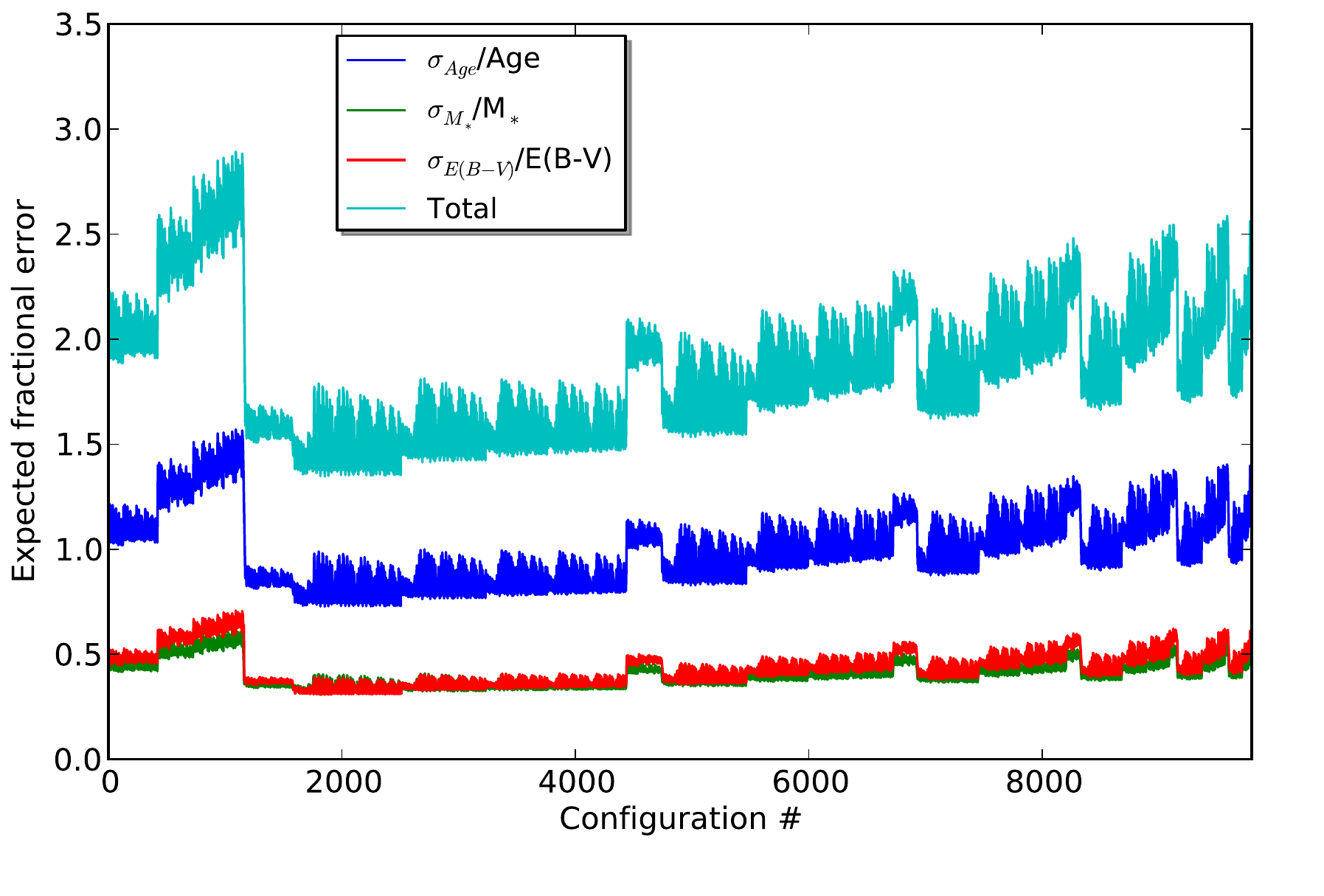}
 \hspace{0.5cm}
  \includegraphics[height=5.5cm]{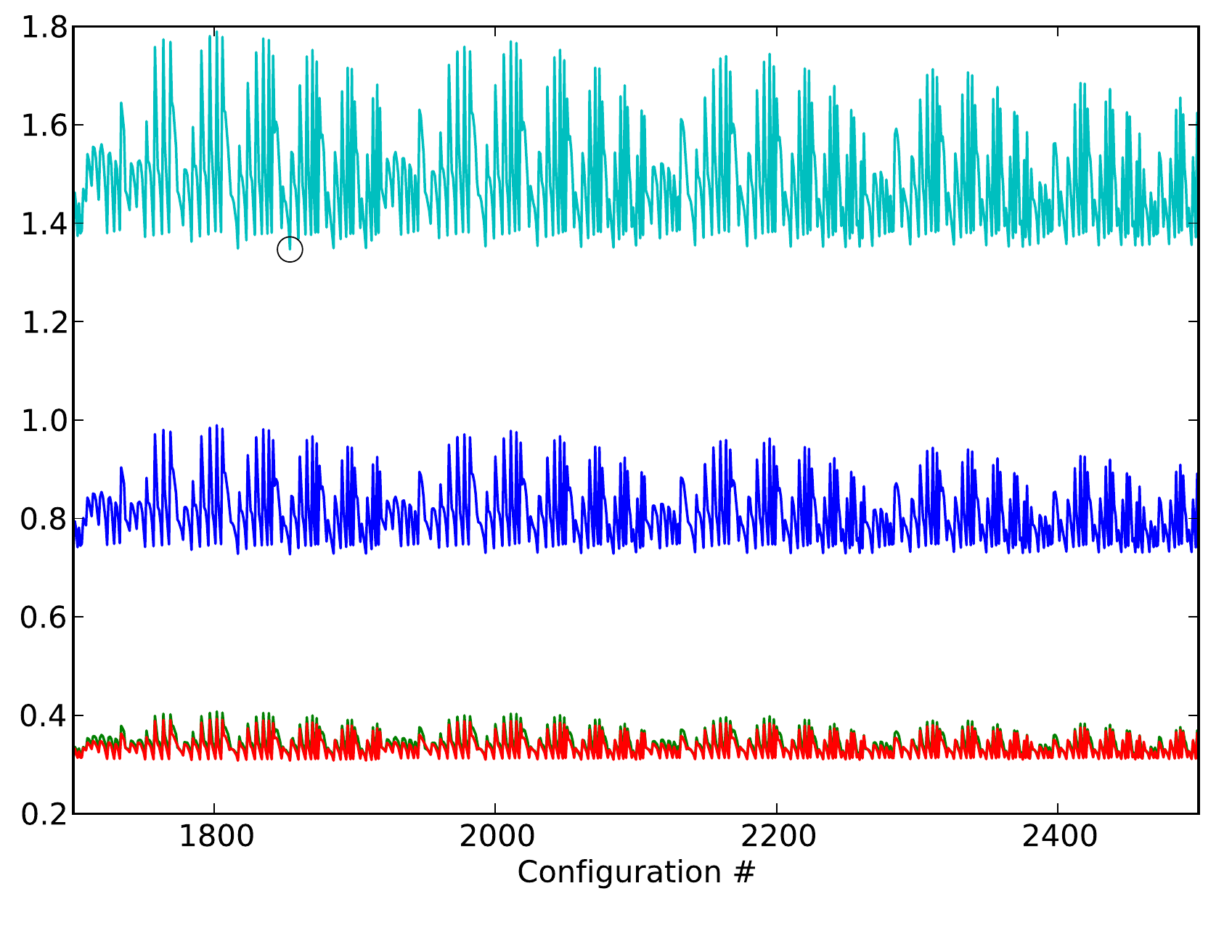}
 \caption{{\bf Optimization of planned HST observation, I.} {\it Left:} The forecasted fractional errors for each of the SED fitting parameter and their total for the 9794 configurations considered in the worked example described in Sec. VI. {\it Right:} The inset shows the small-scales fluctuations of the errors in the region where they are minimal; the circle shows the location of the absolute minimum of the sum of the fractional errors, whose configuration is described in the text.} 
\label{Fig:Example}
\end{figure*}

The results for the uncertainties in the stellar mass are similar to those obtained for the age. In this case, the average relative difference $(\sigma_{\rm MCMC} - \sigma_{\rm FM})/\sigma_{\rm FM}$ is 25\%, the fraction of objects for which the absolute value of the relative difference is less than 0.5 or less than 1.0 is 63 and 85$\%$ respectively, and the median value of the relative difference is again $-5\%$. The trend is exactly equivalent to what is found for the age, as expected since there is a well-known positive correlation in the increasing age vs increasing mass direction for constant star formation. Finally, the dust content parameterized by E(B-V) is the parameter for which the FM performs best; the average relative difference between the uncertainties found by the two methods is 10$\%$, the median disagreement is $1.5\%$, and the fraction of objects for which the absolute value of the relative difference is less than 0.5 or less than 1.0 is 75$\%$ and 96$\%$ respectively. These statistics are also reported in Table \ref{Table:FMGalMC}.
 
 We also investigate whether the capability of GalFish to match the uncertainties obtained via MCMC is correlated with some properties of the target galaxies. In particular, in Fig. \ref{Fig:FMGalMC} we show the relative differences as a function of the $\chi^2$ of the fit, of the signal-to-noise ratio of the SED, defined as $\left[ \sum_{i = 1, n_{\rm bands}} \left(\frac{\phi_i}{\sigma_i}\right)^2\right]^{0.5}$, and of the star formation rate of the galaxy, but we do not observe any significant trend. Further tests done on the correlations of each parameter with the difference in its own uncertainty similarly do not reveal any recognizable correlation.

 \section{Survey planning with GalFish: a worked example}
 \label{secV}
  
  \begin{figure*}
  \centering
 \includegraphics[height=5.5cm]{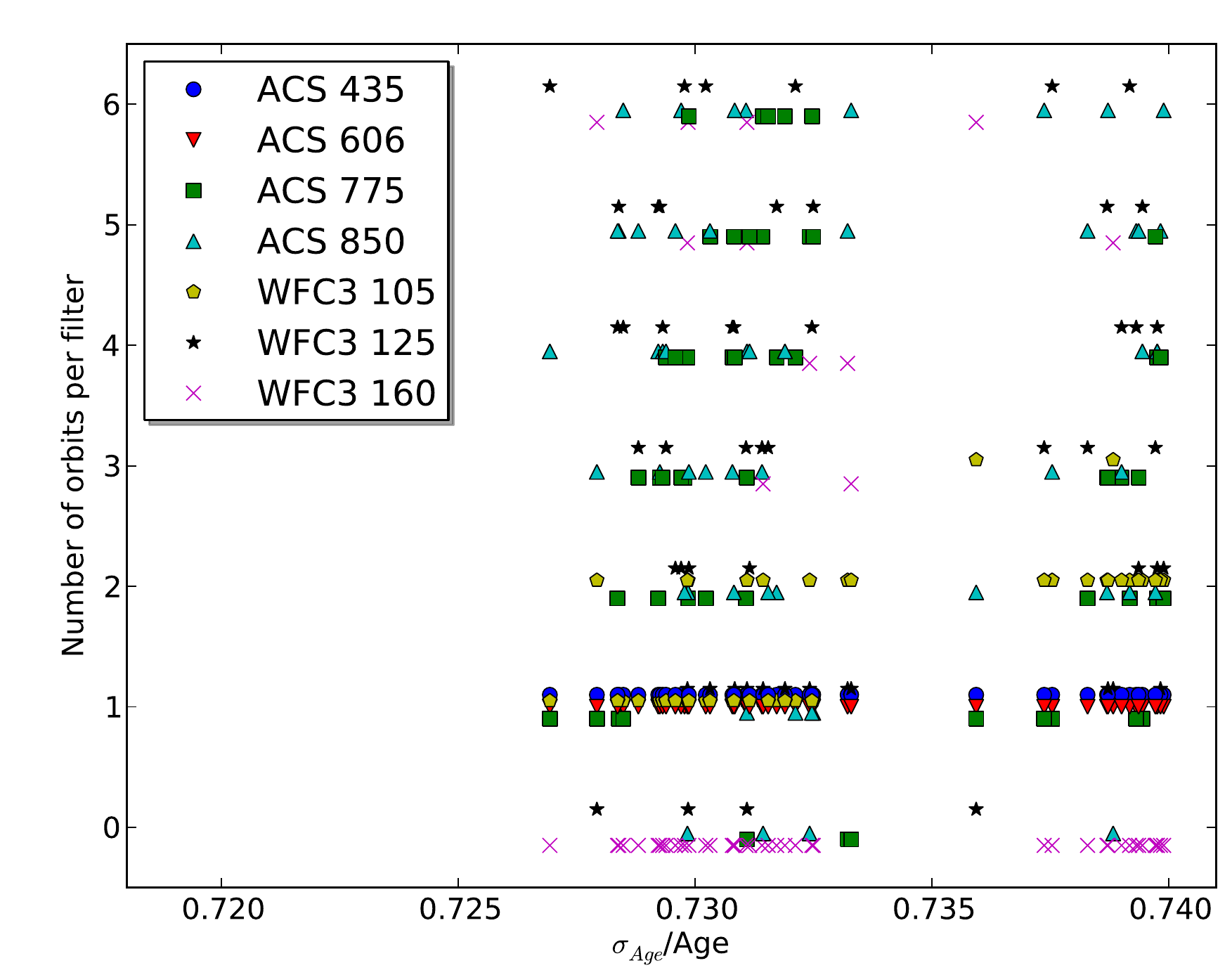} 
  \includegraphics[height=5.5cm]{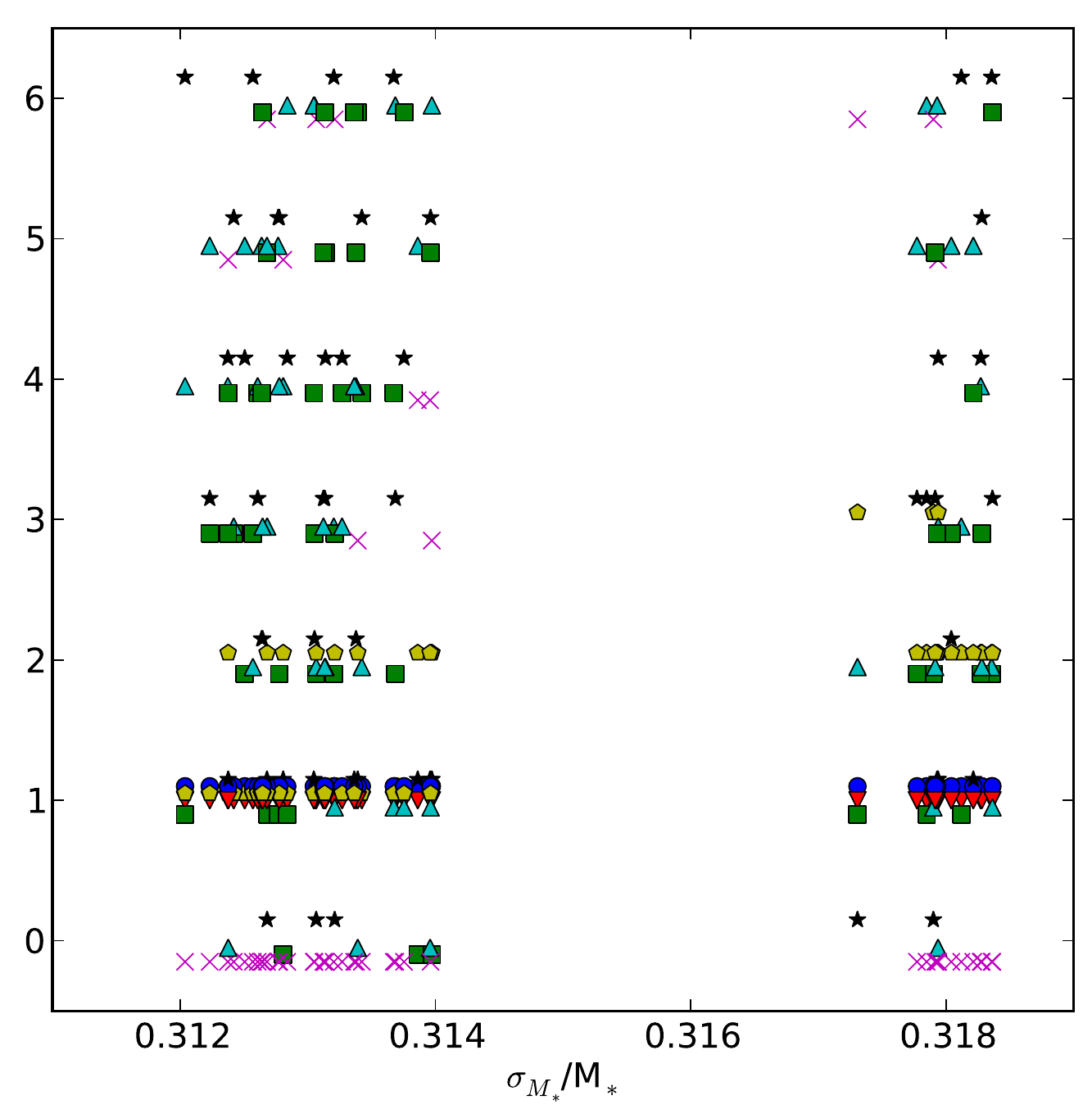} 
   \includegraphics[height=5.5cm]{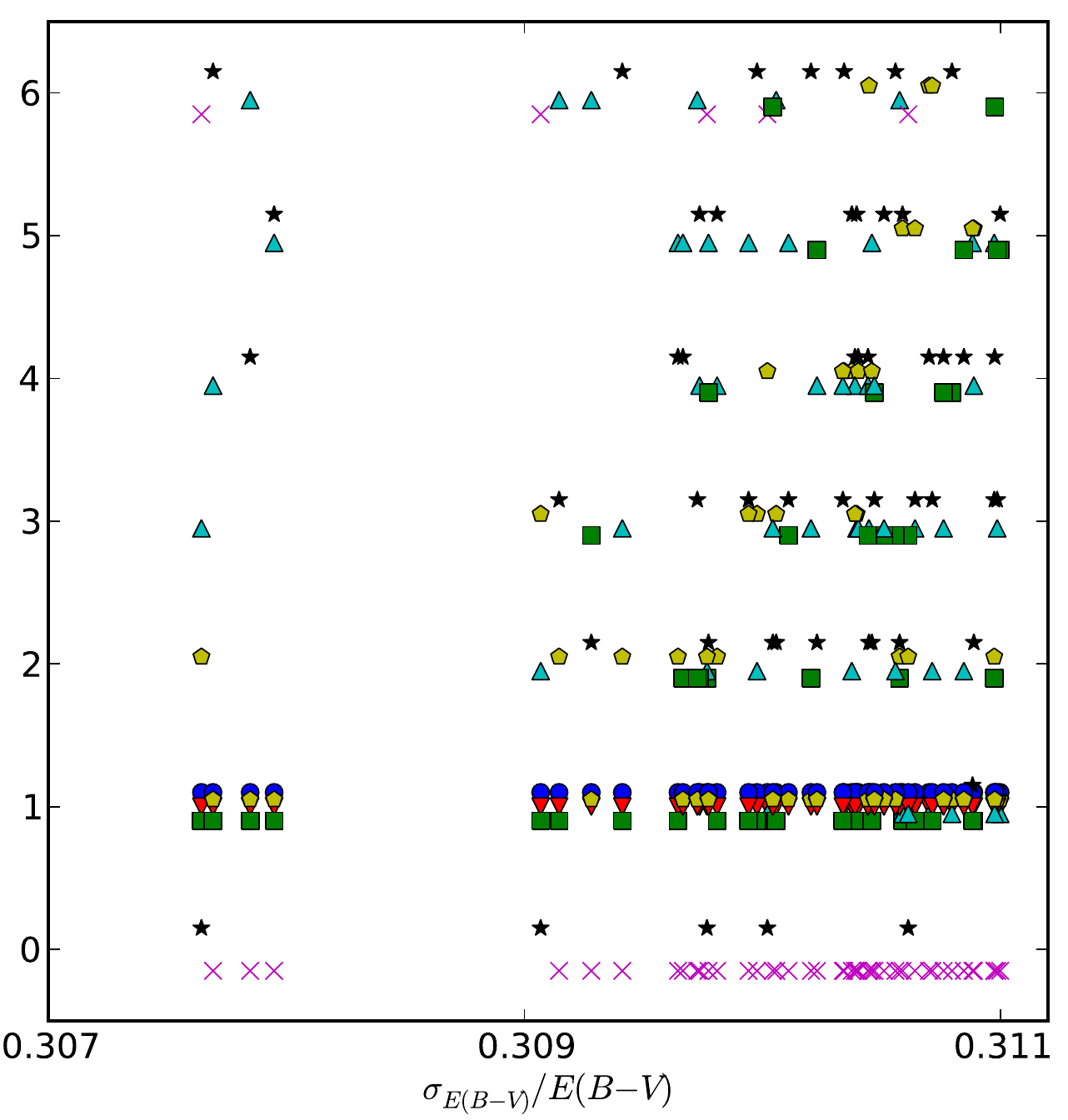} 
\caption{{\bf Optimization of planned HST observation, II.} The three panels show the distribution of time, in orbits, allotted to each of the seven filters for the 50 configurations that minimize the error the SED fitting parameters. In all cases, the error is lower when one orbit is spent in each of the B (ACS, 435W) and V (ACS, 606W) filters, and a time between one and three orbits is spent in the Y (WFC3, 105) filter, while various time allocations for all the other bands give rise to similar errors.}
\label{Fig:Config}
\end{figure*}

In this section we give a practical example of how to optimize observations using GalFish. We assume that Spitzer-IRAC observations at the depths of GOODS-S are available for our target galaxies; we fix the depth of these channels to the ones described in \cite{Dahlen2010} since limited improvements in these measurements will be available in the near future. Our scope is to plan HST observations in the B, V, I, and z channels on the Advanced Camera for Surveys (ACS) and the Y, J and H channels on the Wide Field Camera 3 (WFC3), with a total of 14 requested orbits. Overhead times are not included in this total time. We require that at least five out of these seven filters must be used, that an integer number of orbits is spent in each filter, and that a maximum of six orbits are spent in one single filter; these requirements are merely dictated by common sense and serve to limit the total number of possibilities while preserving the freedom intrinsic in this optimization problem. Each optimization process is valid for a certain type of target galaxy; in this example, we consider a low mass (M$_*$ $\sim 10^9$ M$_\odot$), 1 Gyr old, moderately dusty (E(B-V) = 0.2)  galaxy at $z = 2$, and we run GalFish on each of the 9794 possible observational configurations. In Fig. \ref{Fig:Example}, we show the predicted fractional uncertainties on age, mass and dust content, as well as their sum, which might be used as the overall quantity to minimize. When this indicator is used, the favored configuration is one where one orbit is allotted to the B, V, and I, and Y filters, four orbits are spent in the z filter, six orbits are allotted to the J filter, and the H filter is not used. The resulting best-case uncertainties achievable on each parameter are $\Delta$Age/Age = 70\% or $\Delta$Age = 0.7 Gyr, $\Delta$M$_*$/M$_*$ = 30\% or $ \Delta$M$_*$ = 3$\times10^8$ M$_\odot$, and $\Delta$E(B-V)/E(B-V) = 30\%, or $\Delta$E(B-V) = 0.06. It is interesting to note that the configurations that lead to the lowest fractional error in each parameter are roughly the same as the ones that give the lowest sum; in particular, the aforementioned setup is also optimal to minimize the uncertainty on age and mass, and very slightly sub-optimal (with a difference at the percent level) to minimize the uncertainty on E(B-V).

For all parameters, we observe a ``large scale" modulation which is set by the amount of time spent in the ACS B and V filters; better results are always obtained when one and only one orbit of the total 14 is dedicated to each of these two filters. The only other constant requirement observed in the minimal-errors region is that the time spent observing with the WFC3 Y filter is between one and three orbits. After these variables have been fixed, several hundred configurations generate similar predicted results; the 50 configurations generating the lowest error for each parameter are shown in Fig. \ref{Fig:Config}. 

\section{Discussion}
\label{secVI}
In this paper we have compared the uncertainties on SED fitting obtained by using two different methods, the Fisher Matrix formalism and a MCMC algorithm for PDF fitting, GalMC. In the first case, the forecasted uncertainties come from a mathematical approximation of the posterior distribution function at some fiducial value of the parameters; no data are involved. In the second case, the data are used to properly reconstruct the PDF by sampling the parameter space through a Markov chain. The latter method is much slower but more accurate; in particular, the width of the PDF predicted by the FM is in general predicted to be smaller than the real precision attainable on the parameters measured by data fitting techniques such as MCMC. However, if it can be verified that the uncertainties obtained via the two methods are comparable, the much faster FM becomes a very useful tool to predict the results of a certain experimental setup and to plan astronomical observations.

We have compared the PDFs predicted by the FM and fitted by GalMC for several sets of reference galaxies, both on simulated data and real data. The comparison that uses simulated data is the most relevant one for the purpose of comparing the two methods in the context of planning observations, since in any case a fiducial model needs to be chosen, whether the FM or MCMC is used to compute the uncertainties on the SED fitting parameters. On the other hand, performing the comparison on real data allows one to test whether the simplifying assumptions used in modeling the stellar populations invalidate the attempt itself to predict the achievable precision in the measurement of parameters. 

In the case of simulated SEDs, we started by considering high-mass galaxies  ($M_* \simeq 9 \times 10^9 M_\odot$), but most of our analyses have been conducted on low-mass galaxies ($M_* \simeq 1.5 \times 10^9 M_\odot$), for which more potential sources of disagreement are present as a consequence of the lower signal to noise ratio. We considered dusty and dust-free galaxies, with constant and instantaneous star formation history, with and without infrared photometric coverage, at redshifts z = 1, z = 2, and z = 3. We found a remarkably good agreement between the results from the two methods. Of course, the FM is not able to capture features such as non-Gaussianity or multi-modality of the posterior, which MCMC can correctly explore. However, the precision in the SED fitting measurements for the two techniques are within 30\% of each other (with the FM always being an underestimate) in the vast majority of cases, and within a factor of two in all cases. Priors play an important role in the comparison, in particular for higher redshift galaxies with low S/N. MCMC uses a top-hat prior in age, (6 $ <  \log_{10}$(age/yr) $< $ 10.13), bracketed by the minimum age for which templates are trustworthy and by the Universe' age. This prior needs to be incorporated in the FM at least approximately, otherwise the FM may grossly overestimate the uncertainties. We have found that if the fiducial model is far from the extremum values of the age range, a good proxy for the top-hat prior in age  is a Gaussian prior of width $\Delta \ln$(age/yr) $\simeq$ 3.5. As a general prescription, one can always test the predictions with or without the prior in age, but it is easy to verify that the prior only has a non-negligible effect if the prior intersects the distribution predicted by the FM at high likelihood values. We also investigated the effect of the range imposed in E(B-V) (between 0 and 1 in GalMC), finding this prior to be negligible for all cases considered. 

We then proceeded to test the performance of the FM approximation for a set of 485 galaxies at redshifts between 0 and 5 observed in GOODS-S, with widely spread photometric and stellar population properties. In this case, we also found that the FM provides an acceptable approximation to the probability distributions found by GalMC, with average relative differences between results of the order of $10-20\%$, and significant failures (errors differing by more than a factor of two) in $\sim$ 10\% of all cases. As mentioned earlier, these difference do not necessarily arise from the failure of the FM formalism, but can come from the fact that a real galaxy SED might be more complicated than the model used for the analysis, and the fiducial model used to predict the uncertainties might be significantly different from the observed spectrum.

In this analysis, we chose to only marginalize over three SED fitting parameters: age, mass, and dust content. It is often the case that other parameters are considered in SED fitting, for example, the e-folding time $\tau$ for exponential star formation histories, the metallicity $Z$, or the  redshift. As far as $\tau$ is concerned, we showed in \cite{2011ApJ...737...47A} that for exponential SFHs the parameters age and $\tau$ are nearly degenerate, and that different choices of prior on $\tau$ can strongly influence the results of SED fitting, even for those data, which have deeper photometry than the ones considered in this paper. Therefore, $\tau$ is not a suitable parameter for FM analysis. However, an exponentially declining or decreasing SFH with $\tau <<$ age is effectively a starburst either at the age of the galaxy or at very recent times, and an exponentially increasing or declining SFH with $\tau \simeq$ age is well approximated by a constant SFH, so that the results found here apply to most exponential SFH models. Similarly, the SEDs depend on metallicity in a very non-linear manner and thus the effect of $Z$ cannot be effectively captured in the FM formalism. The present analysis assumes that the metallicity of the target galaxies is known, which is a fairly common assumption made in the literature when analyzing broad-band photometry data. However, if one were to relax this hypothesis, the width of the recovered PDFs could increase. Our analysis in \cite{Acquaviva_Evol} showed that this effect is mild in the case of $z \sim 2$ and $z \sim 3$ Lyman Alpha Emitters with UV, optical, NIR and IRAC photometry, with probability distributions becoming $\sim$ 10-20\% broader, but of course this effect might be stronger for different stellar populations and data sets and this caveat should be kept in mind.

In the case of redshift, a direct comparison with GalMC is not really informative, since for real data the redshift is usually determined by using photo-z algorithms. We have been so far very cautious in using GalMC to determine photometric redshifts. In fact, when the PDF has multiple and widely separated peaks, even MCMC codes might fail to explore the parameter space in a way that respects the basic MCMC property that the density of visited points is proportional to the volume enclosed in the probability distribution function. As a result, the analysis presented here is strictly valid only if spectroscopic redshifts are available. However, the FM formalism is able to include redshift as one of the SED fitting parameters, at least in the particular form of a Gaussian prior centered at the photometric redshift of the galaxy. If the uncertainties on the photo-$z$ can be expressed in this form, and an educated guess of the width of such prior is available, we can expect the uncertainties from the FM to be reliable.
 
In summary, we have demonstrated that the Fisher Matrix formalism is a valuable tool for planning photometric observations. Estimates of the attainable precision on SED fitting parameters are relatively accurate (within a factor of two at the maximum) and can be obtained very quickly, allowing one to optimize planned surveys by simulating a variety of experimental setups. Just as one example of the many applications, suppose that the scientific goal of an experiment is to achieve a certain precision in the measurement of stellar mass. The FM formalism allows one to explore what wavelength coverage and depth of observations is needed, and to understand how the aim can be achieved by minimizing costs.

In the hope that the formalism cast here is useful to the scientific community, the Fisher Matrix code presented and used here is publicly available as a PHP website at http://galfish.physics.rutgers.edu. Users can specify which filters they want to use from a list of many common ones, the depth of planned observations, and the characteristics of the target galaxies, and they will immediately obtain the results of the FM analysis. The Fortran code used in the more complex optimization problems will also be available on request upon publication of the paper. \\

\acknowledgements

We warmly thank the anonymous referee of \cite{2011ApJ...737...47A} for her/his invitation to compare the results obtained by GalMC to the predictions from the Fisher Matrix formalism, Tomas Dahlen for providing us with the ACS-z selected  TFIT catalog used in Sec. V, and Lucia Guaita, Amir Hajian, and Licia Verde for useful comments on the manuscript draft. This material is based on work supported by the National Science Foundation under grants AST-0807570 and AST-1055919. Support for Program number HST-GO-12060.57 was provided by NASA through a grant from the Space Telescope Science Institute, which is operated by the Association of Universities for Research in Astronomy, Incorporated, under NASA contract NAS5-26555.
  
 \begin{table*}[t!]
  \center
  {\footnotesize \resizebox{\textwidth}{!}{
  \begin{tabular}{@{\extracolsep{\fill}}lccccccccc}
  
    \hline
    \hline
\small     & F435W  &  F606W  &  F775W     & F850LP   &  F105W  & F125W & F160W   & IRAC I	  & IRAC II \\

    \hline
     \hline
Central $\lambda \; [\AA]$ & 4350 & 6060 & 7750 & 8500 & 10500 & 12500 & 16000 & 35634 & 45110 \\
High Mass, z  = 1  & 0.37 & 0.5 & 1.0 & 1.6 & 2.0 & 2.6 & 3.6 & 9.9 & 8.5 \\
High Mass, z  = 2   & 0.023 & 0.098 & 0.086 & 0.13 & 0.17 & 0.38 & 0.51 & 1.8 & 2.3\\
High Mass, z  = 3   & 0.001 & 0.0078 & 0.043 & 0.11 & 0.074 & 0.077 & 0.16 & 0.59 & 0.89 \\
Low Mass, z  = 1 & 0.06 & 0.086 & 0.17 & 0.27 & 0.33 & 0.42 & 0.59 & 1.65 & 1.42  \\
Low Mass, z = 2 & 0.009 & 0.014 & 0.018 & 0.023 & 0.033 & 0.059 & 0.084 & 0.31 & 0.38\\
Low Mass, z = 3 &  0.0014 & 0.0045 & 0.0069 & 0.0082 & 0.0095 & 0.012 & 0.022 & 0.1 & 0.14\\
Dust-free, z = 1 & 0.64 & 0.6 & 0.78 & 0.98 & 1.1 & 1.2 & 1.2 & 2.0 & 1.5\\
Dust-free, z = 2 & 0.14 & 0.18 & 0.17 & 0.13 & 0.19 & 0.27 & 0.34 & 0.39 & 0.51 \\
Dust-free, z = 3 & 0.038 & 0.078 & 0.061 & 0.064 & 0.092 & 0.05 & 0.076 & 0.19 & 0.26 \\
Starburst, z = 1 & -0.015 & 0.0065 & 0.023 & 0.051 & 0.12 & 0.11 & 0.32 & 0.9 & 0.81 \\ 
Starburst, z = 2 & 0.0011 & 0.0047 & 0.0017 & -0.0002 & -0.017 & 0.014 & 0.034 & 0.17 & 0.16 \\
Starburst, z = 3 & 0.013 & 0.024 & 0.023 & -0.066 & 0.013 & -0.053 & -0.065 & 0.066 & 0.029 \\
No IRAC, z = 1 & 0.06 & 0.086 & 0.17 & 0.27 & 0.33 & 0.42 & 0.59 & - & - \\
No IRAC, z = 2 & 0.009 & 0.014 & 0.018 & 0.023 & 0.033 & 0.059 & 0.084 & - & - \\
No IRAC, z = 3 &  0.0014 & 0.0045 & 0.0069 & 0.0082 & 0.0095 & 0.012 & 0.022 & - & - \\
$5\sigma$ limit [mag] & 26.4 & 26.8 & 26.2 & 25.6 & 26.2 & 25.6 & 25.6 & 26.0 & 25.4 \\
$1\sigma$ flux error [$\mu$Jy] & 0.02 & 0.014 & 0.024 & 0.042 & 0.024 & 0.024 & 0.042 & 0.029 & 0.05\\

     \hline    
     \hline
  \end{tabular}
  }}
\caption{\label{table:mag} Fluxes, $5\sigma$ limiting magnitudes, and $1\sigma$ errors in $\mu$Jy used for SED fitting on simulated galaxies with GalMC and for Fisher Matrix forecasts. The simulated fluxes include a photometric Gaussian scatter of amplitude equal to the error.}
\end{table*}

\begin{table*}[h!]
  \center
  \vspace{-14cm}
  {\footnotesize \resizebox{\textwidth}{!}{
  \begin{tabular}{@{\extracolsep{\fill}}lccccc}
    \hline
    \hline
\small    Quantity & \small Average & \small Median & \small Standard & \small Fraction of objects & \small Fraction of objects \\
\small & & & \small Deviation & \small  for which $ \frac{|\sigma_{MCMC} - \sigma_{FM}|}{\sigma_{FM}} <  0.5$ & \small for which  $\frac{|\sigma_{MCMC} - \sigma_{FM}|}{\sigma_{FM}} <  1.0$ \\
    \hline
     \hline

    $\frac{(\sigma_{MCMC} - \sigma_{FM})}{\sigma_{FM}}$ , \small Age  & 0.1 & -0.05 & 0.63 & 68\% & 91\% \\   
     $\frac{(\sigma_{MCMC} - \sigma_{FM})}{\sigma_{FM}}$  , \small M$_*$  & 0.26 & -0.05 & 0.99 & 63\% & 85\% \\  
      $\frac{(\sigma_{MCMC} - \sigma_{FM})}{\sigma_{FM}}$ , \small E(B-V)      & 0.1 & 0.015 & 0.5 & 75\% & 96\% \\  
        \hline    
     \hline
  \end{tabular}
  }}
\caption{The relative difference  $\frac{(\sigma_{MCMC} - \sigma_{FM})}{\sigma_{FM}}$  in the uncertainty on each SED fitting parameter found by the Fisher Matrix and by GalMC describes the disagreement between these two methods. Here we summarize the statistical properties of the distribution of the differences shown in Fig. \ref{Fig:FMGalMC}, which refer to a set of 485 galaxies observed in GOODS-S.}
\label{Table:FMGalMC} 
\end{table*}

\end{document}